\documentclass[useAMS,usenatbib,usegraphicx]{mn2e}
\usepackage{color}
\usepackage{graphicx}
\usepackage{rotating}
\usepackage{amssymb}

\def \apj {ApJ}
\def \aj {AJ}
\def \apjl {ApJ Lett}
\def \apjs {ApJS}

\def \mnras {MNRAS}

\def \pasp {PASP}

\def \aap {A\&A}
\def \araa {Annual Review of A\&A}

\def \nar {New Astronomy Reviews}

\title[Near-infrared counterpart of faint transients]{A search for near-infrared 
counterparts  of two faint neutron star X-ray transients  : XMMU J174716.1--281048 and SAX J1806.5--2215}
\author[Kaur et al.]{Ramanpreet Kaur$^{1}$\thanks{Email : raman.pk@gmail.com}, 
Rudy Wijnands$^{2}$,
Atish Kamble$^{3}$, 
Edward M. Cackett$^{4}$, 
\newauthor 
Ralf Kutulla$^{5}$, 
David Kaplan$^{5}$, 
Nathalie Degenaar$^{2,6}$\\
$^1$ Physics Department, Suffolk University, 41 temple street, Boston, Massachusetts, 02114, USA \\
$^2$ Anton Pannekoek Institute for Astronomy, University of Amsterdam, 
Science Park 904, 1098 XH, Amsterdam, The Netherlands \\
$^3$ Harvard-Smithsonian Center for Astrophysics, 60 Garden Street, Cambridge, MA 02138 \\
$^4$ Department of Physics and Astronomy, Wayne State University, Detroit, MI 48201, USA\\ 
 $^5$ Physics Department, University of Wisconsin-Milwaukee, Milwaukee, WI 53211, USA\\
   $^6$ Institute of Astronomy, University of Cambridge, Madingley Road, Cambridge CB3 OHA, UK}

\begin{document}

\date{\today}

\pagerange{\pageref{firstpage}--\pageref{lastpage}} \pubyear{2016}

\maketitle

\label{firstpage}

\begin{abstract}
We present our near-infrared (NIR) imaging observations of  two neutron star low mass X-ray binaries XMMU J174716.1--281048 
and SAX J1806.5--2215 obtained using the PANIC instrument on the 6.5-meter Magellan telescope and 
 the WHIRC instrument on the 3.5-meter WIYN telescope respectively. Both sources are members of the class of faint to 
very-faint X-ray binaries and undergo very long X-ray outburst, hence classified as `quasi persistent X-ray binaries'.  
While XMMU J174716.1--281048 was active for almost 12 years between 2003 and 2015,  
SAX J1806.5--2215 has been active for more than 5 years now since 2011.
From our observations, we identify two NIR stars  consistent with the  {\it Chandra} X-ray error circle of  
XMMU J174716.1--281048. The comparison of our observations with the UKIRT Galactic plane observations 
taken during the same outburst, color-color diagram analysis and spectral energy distribution
suggest that both stars are probably a part of the field population and are likely high mass stars. 
Hence possibly neither of the two stars is a true NIR counterpart.  
For the faint X-ray binary SAX J1806.5--2215 during its current outburst, we detected a NIR star in
our K band WIYN observations consistent with its {\it Chandra} error circle. 
The identified NIR star was not detected during the UKIRT observations taken 
during its quiescent state. 
The comparison of two observations suggest that there was an increase in flux by at least one magnitude 
of the detected star during our observations, hence suggests  the detection of the 
likely counterpart of SAX J1806.5--2215. 
\end{abstract}

\begin{keywords}
 binaries: general,  X-rays: binaries, stars: neutron, infrared : stars, accretion, accretion disks, stars: individual (XMMU J174716.1--281048, SAX J1806.5--2215)

\end{keywords}

\section{Introduction}

Low-mass X-ray binaries (LMXBs) are binary systems with a compact star (neutron star or black-hole) accreting matter 
from a low-mass star (usually K, M dwarf or red giant,  sometimes a white dwarf) 
while they revolve around each 
other. Many of these systems remain dim for most of their time and 
occasionally undergo a large increase in X-ray luminosity  (by a factor $>$ 1000) up to 
10$^{37-39}$ erg  s$^{-1}$ (2 - 10 keV). These X-ray  outbursts 
last for a few weeks to months except in some of the LMXBs in which they last up to a few years or on rare 
occasions, decades.  
The sources with these long X-ray outbursts ($>$ 1 year) are also known as  quasi-persistent LMXBs.
During these outbursts,  the majority of the X-rays are emitted in the inner accretion disk while the 
optical/infrared (OIR) light is typically emitted due to the reprocessed emission  in  the outer accretion disk and 
the surface of the companion star. 
In the quiescence state, the OIR light is often dominantly emitted from the low-mass companion star unless the 
companion star is a very low mass star or a white dwarf star. In such case, the disk might still contribute 
significantly (or even dominate) the OIR light.
This change in the OIR light
allows us to pin-point the type of  companion star (or get stringent upper limits) 
which is important to constrain binary parameters of the object (i.e., the  
mass of the donor and the compact object, orbital period).  

During the last decade, many faint to very-faint X-ray transients have been discovered which 
have peak X-ray luminosities of 
10$^{34-36}$ erg s$^{-1}$ (e.g., \citealt{Wijnands2006AA}, \citealt{Degenaar2010AA}). 
Most of these sources are discovered\ in the Galactic plane, with the highest concentration 
near the Galactic centre. A large fraction of them are discovered through  
thermonuclear bursts (\citealt{Cornelisse2002AA392}, \citealt{Degenaar2011}, \citealt{Wijnands2009}, 
\citealt{Campana2009ApJ}). These bursts have so far been observed only 
from the neutron star LMXBs hence probably these faint transients are also such kind of systems. 
However the current instability models that explain the transient behaviour of LMXBs  
(e..g, \citealt{Lasota2001}) cannot explain the low peak X-ray luminosities of these faint systems. 
It is possible that these sources are accreting from the stellar wind of a companion star 
(\citealt{Pfahl2002ApJ}, \citealt{Degenaar2010AA}, \citealt{Maccarone2013}) or they 
have a small accretion disk, which indicates  a 
small orbit that can only accommodate a very small donor star - e.g., brown  or 
white dwarf \citep{King2006}. The latter are called ultra-compact 
X-ray binaries with extremely short orbital periods ($<$ 80 minutes; 
\citealt{Nelemans2010}, \citealt{IntZand2005}, \citealt{IntZand2007}, \citealt{Hameury2016}). 
In an alternative interpretation, it has been proposed
 that the magnetic field of the neutron star is inhibiting accretion in such sources, resulting in very faint 
 X-ray luminosities
(e.g., \citealt{Heinke2009}, \citealt{DAngelo2012},
\citealt{Heinke2015}, \citealt{Arnason2015}). 

In order to investigate the above proposed scenarios, we must find the nature of the companion star. 
However, as most of these sources are present near the Galactic centre, their detection in OIR is 
largely hampered by the high extinction and the crowded fields. It is important to 
find the position of these X-ray sources accurately with the smallest error circle possible, to  
locate the OIR counterpart. When an OIR counterpart has been identified, the exact identification of 
the companion star could  be done through the OIR spectroscopy. 

Not much has been reported in the OIR regime of very faint  X-ray binaries so far. \cite{Degenaar2010MNRAS} detected 
the OIR counterpart of the neutron star binary 1RXH J173523.7--354013 with a magnitude of 15.9 mag in optical R band. 
The follow up optical spectroscopic observations of the source
showed a narrow H$_{\alpha}$ emission line, indicating 
a relatively normal low mass X-ray binary and ruling out an ultra compact X-ray binary.
The OIR counterpart of a globular cluster faint X-ray transient M15 X-3 was detected by \cite{Heinke2009} using  
Hubble space telescope observations, with a magnitude of 22 mag in V band.  
Later on,  \cite{Arnason2015} used additional more recent Hubble Space Telescope observations and 
estimated the companion star to be a  main sequence star of mass  0.44 M$_\odot$ in a $\sim$ 4 hour orbital period. 
The multiple optical observations  of the faint black hole X-ray binary Swift J1357.2--0933 detected the source
at a quiescent R band magnitude of $\sim$ 22.2 mag. The companion star in this binary has been estimated to be 
a M4 type main sequence star in a 2.4 hour orbit around the black hole 
(\citealt{Rau2011}, \citealt{Corral2013}, \citealt{Armas2014}). In all these cases, these very faint 
systems turned out to be 
ordinary low mass X-ray binaries and not ultra compact X-ray binaries. Hence it is possible that the magnetic field of 
the compact object is inhibiting  accretion in all these cases as proposed by  \cite{DAngelo2012} and 
\cite{Heinke2015}. 
To say anything more certain we must find the nature of companion star in more faint X-ray binaries. At this point, we 
cannot exclude that we might be looking at multiple classes of sources in the faint regime but these sources are definitely bound to open
 new windows  on binary evolution. 

In this paper we present our near-infrared (NIR) imaging observations of two 
faint to very faint quasi-persistent transients to search for their counterparts.

\subsection{XMMU J174716.1--281048}

XMMU J174716.1--281048 (hereafter XMM1747) was serendipitously discovered during  a pointed 
{\it XMM-Newton} observation of the supernova remnant G0.9+0.1 performed on March 12, 2003. The source had an X-ray 
 flux of 3.7 $\times$ 10$^{-12}$ erg cm$^{-2}$ s$^{-1}$ in 2-10 keV energy band (\citealt{Sidoli2003ATel147},  
\citealt{Sidoli2004}, \citealt{DelSanto2007}). 
The source X-ray spectrum  was best fitted with an absorbed powerlaw model with photon  index $\Gamma$ of 2.1 
and a hydrogen column density $\it N_{\rm H}$  of 8.9  $\times$ 10$^{22}$ cm$^{-2}$. At a distance of 8.4 kpc 
(inferred from a type-I X-ray burst analysis; \citealt{Degenaar2011}, see below for more information about the 
bursts seen in this source), the corresponding X-ray luminosity of the source is  
$\sim$ 3 $\times$ 10$^{34}$  erg s$^{-1}$. The source 
was not visible in the previous {\it XMM-Newton} and {\it Chandra} observations performed 
between September 2000 and
July 2001 which implied that the quiescent 
X-ray luminosity of the source was $\le$ 10$^{32}$ erg s$^{-1}$  \citep{DelSanto2007}.
The rise in flux by a factor more than 100 and  the low peak X-ray luminosity classifies the source as a  
very faint X-ray transient \citep{DelSanto2007}. 

On March 22, 2005, {\it INTEGRAL}/JEM-X  detected a transient IGR J17464--2811 because it exhibited a type-I 
X-ray burst \citep{Brandt2006ATel970}. 
The position coincidence of this burst source  with the position of XMM1747,   
which was also active during observations made on Feb 25-26, 2005  using the {\it XMM-Newton} satellite \citep{Sidoli2006}
indicated that both  sources are the same source \citep{Wijnands2006ATel972}. This classifies XMM1747 
as a neutron star low mass X-ray binary. 
On the basis of the properties of the X-ray burst observed by  {\it INTEGRAL}, it was suggested that the
source was likely continuously active between 2003 and 2005 and was undergoing a 
long-term outburst  \citep{DelSanto2007}. Another type-I X-ray burst was observed from this source on August 13, 
2010 \citep{Degenaar2011} using the {\it Swift}/Burst Alert Telescope (BAT).  
Since 2007, XMM1747 has been observed several times using the 
{\it Swift} satellite and once using the {\it Chandra} satellite. The source 
has been detected at  similar flux levels and with similar spectral parameters from all the observations taken until 
March 2014 
(\citealt{Degenaar2007ATel1078},  
\citealt{Degenaar2007ATel1136}, \citealt{Campana2009}, 
\citealt{Sidoli2007ATel1174}, \citealt{DelSanto2007ATel1207}, \citeyear{DelSanto2007},
\citeyear{DelSanto2009ATel2050},  
\citeyear{DelSanto2010ATel2624}, 
\citeyear{DelSanto2011ATel3471}, 
\citeyear{DelSanto2012ATel4099},
\citeyear{DelSanto2015ATel7293}).

The last set of observations was made using the {\it Swift} satellite in 
March 2015 during which the source was not detected with a 3-sigma upper limit on the luminosity of 
7 $\times$ 10$^{33}$ erg s$^{-1}$ (2 - 10 keV)
at 8 kpc (\citealt{DelSanto2015ATel7293}), which indicates that the source 
 went down considerably and possibly to quiescence although the latter statement need to be confirmed 
with more sensitive {\it Chandra} or {\it XMM} observations.

The best known position of the source so far is RA : 17$^\mathrm{h}$  47$^\mathrm{m}$ 16$\fs$16, 
Dec. : $-28^{\circ}$ $10^{\prime}$ $48${\farcs}$0$ with an  error circle of $0\farcs5$ \citep{Degenaar2007ATel1136}. 
A possible NIR counterpart of XMM1747 was reported  \citep{Degenaar2007ATel1136} at this position from the observations 
obtained on  May 27, 2007 using 
the instrument ANDICAM  on the 1.3-meter SMARTS telescope. The source had a {\it H} band magnitude of 
15.3 $\pm$ 0.1 mag but was 
not detected in the {\it V} and  {\it I} band.

\subsection{SAX J1806.5--2215}

SAX J1806.5--2215 (hereafter SAX1806) was discovered using the {\it BeppoSAX}/Wide Field Cameras (WFC) 
through the detection of two type-I X-ray bursts observed between August 1996 to October 1997 
(\citealt{In'tzand1999}, \citealt{Cornelisse2002}). 
At the time when  these X-ray bursts occurred, no persistent X-ray emission was 
detected from the source. However, it was later found using the  {\it RXTE}/All Sky Monitor 
data of the source that it was faintly active for at least nearly two years at this time (from early 1996 till late 1997) with a peak persistent 
flux of  2 $\times$ 10$^{-10}$ erg cm$^{-2}$ s$^{-1}$ (2-10 keV; \citealt{Cornelisse2002}). 
Using the unabsorbed bolometric peak flux of the brightest  burst, 
an upper limit of 8 kpc on source distance was obtained  \citep{Cornelisse2002}. Using this distance, the outburst accretion 
X-ray luminosity of the source was  2 $\times$ 10$^{36}$ erg s$^{-1}$.  {\it Chandra} and {\it Swift} observations of the source  obtained  between 2000-2009 did not detect the source 
with an upper limit on the X-ray luminosity of (0.5 - 2) $\times$ 10$^{33}$ erg s$^{-1}$ 
(\citealt{Chakrabarty2011ATel3218}, \citealt{Degenaar2011ATel3202}, \citealt{Campana2009}).

On February 22, 2011, after 12 years in quiescence,  SAX1806 displayed another X-ray outburst \citep{Altamirano2011}.
{\it Swift} observations obtained on March 1, 2011 
detected  the source within the {\it BeppoSAX} error circle of the X-ray source. The
X-ray spectrum from these observations could be well fitted with a powerlaw model of photon index 
$\Gamma$  
of 2 and N$_H$  of  5.6 $\times$ 10$^{22}$ cm$^{-2}$ \citep{Degenaar2011ATel3202}. 
The unabsorbed flux of the source suggest an X-ray luminosity of  2 $\times$ 10$^{36}$ erg s$^{-1}$ 
for a  distance of 8.0 
kpc  \citep{Degenaar2011ATel3202} which is similar to the X-ray activity level seen during its previous outburst.  
Since the beginning of the outburst in February 2011, SAX1806 has been observed several times 
using the {\it Swift} satellite and once using the {\it Chandra}  satellite. The observations taken till October 2015
reported SAX1806 at  similar fluxes  (\citealt{Kaur2012ATel3926}, \citealt{DelSanto2012ATel4017}, \citealt{VSguera2015ATel8222}). 
Hence we can conclude that the source has been active for $> 5$ years till now.  
The {\it Chandra} observations provided the best known position of the source with a 0$\farcs$6 accuracy (90\% confidence level) at
RA : 18$^\mathrm{h}$ 06$^\mathrm{m}$ 32$\fs$177 and 
Dec : -22$^\circ$  14$^{\prime}$  17$\farcs$20   \citep{Chakrabarty2011ATel3218}. 
The low peak X-ray luminosity of the source and the long outburst classify SAX1806  
as a faint quasi-persistent X-ray transient. 

During the X-ray outburst, a NIR counterpart was detected in our $K_{\mathrm{s}}$ band observations obtained in March 2011 
using the WHIRC instrument on the 3.5-meter WIYN telescope \citep{Kaur2011ATel3268} and the 
observations will be discussed in detail in this paper. 
 
\section{Data reduction and analysis}

XMM1747 was observed on May 25, 2008 in  the $J$,  $H$ and $K_{\mathrm{s}}$ wavebands using the 
PANIC instrument (\citealt{Martini2004}) mounted on the 6.5-meter  Baade Magellan telescope at Las Campanas Observatory in Chile; 
PANIC has a $1024 \times1024$ array of pixels 
with 0$\farcs$125/pixel sampling,  corresponding to  $2^\prime$ $\times$ $2^\prime$ field of view.  
All the science exposures were taken as a set of 5 dithers and a loop of three 10 s exposures were performed 
at each of the 5 dither points (dice 5 pattern). The source was observed for a 
total exposure time of 300 s, 300 s and 750 s in the $J$, $H$ and $K_{\mathrm{s}}$ filters, respectively. 

SAX1806 was observed in the $K_{\mathrm{s}}$ band on March 23, 2011 using the WIYN High-Resolution
Infrared Camera (WHIRC, \citealt{Meixner2010}) mounted on the 3.5-meter WIYN telescope located at  Kitt Peak National Observatory in USA.  
WHIRC has a $2048 \times 2048$ array with $0\farcs1$ arcsec/pixel sampling,
resulting in a field-of-view of $3.3' \times 3.3'$. The source was observed for a total exposure time
of 720 s, which include 40 s exposures at each of the 3 $\times$ 3 dither pattern, repeated twice.

Data reduction for XMM1747 is performed using the PANIC data reduction 
package version 0.95\footnote{http://code.obs.carnegiescience.edu/panic}  (\citealt{Martini2004}). For  
SAX1806, we used a custom-made reduction pipeline  that largely
follows the recipe outlined in the WHIRC Data Reduction 
Guide\footnote{http://www.noao.edu/kpno/manuals/whirc/WHIRC\_Datared\_090824.pdf}. In short, 
for both the sources, after removing the inherent detector nonlinearity, we performed a background
subtraction, flat-fielding, geometric distortion correction, alignment and
finally stacking of science images. The seeing in the stacked XMM1747 images is $0\farcs5$ and  
in the SAX1806  $K_\mathrm{s}$ image is $0\farcs6$.

Flat fielding of the XMM1747 images is performed using the twilight sky flats while for flat fielding of the 
SAX1806 images is performed using the dome flats. 
The dome flats are preferred for WIYN observations due to the 
open structure of the telescope and its susceptibility to stray light from the relatively bright sky. 
Dome flat-fields were taken with the flat-field lamps
switched on and off to account for thermal emission from the warm dome. 
Sky-background levels of science images were estimated by iterative
sigma-clipping of the brightness distribution of each of the
data-frames. Multiple frames at different dither positions were then normalised
by their respective sky-levels, median-combined to remove stars, and then scaled
to the sky-level of each data frame and subtracted. 

We determined  the astrometry for both the sources using the Two Micron All Sky 
Survey (2MASS; \citealt{Skrutskie2006}) 
stars and fitting them for the 
transformation using \texttt{ccmap}\footnote{http://iraf.noao.edu/}.  The solution for the $J$ image of XMM1747  had an rms error
of $0\farcs03$ in each coordinate and we transferred this solution to the {\emph H} 
and $K_\mathrm{s}$ wavebands also. For
SAX1806, the $K_\mathrm{s}$ image had a rms error of  $0\farcs03$ and $0\farcs05$ 
in X and Y coordinate respectively. 

We used standard routines in the {\it IRAF} daophot \citep{Stetson1987} to measure  the instrumental magnitude of stars in our observations.
To calibrate 
the observations, we used the archival 3.5-meter United Kingdom Infra-red Telescope (UKIRT) observations of the source field 
in $J$, $H$ and $K$ bands which were obtained as a part of the UKIDSS Galactic plane survey \citep{Lucas2008}. 
These observations were taken on July 18, 2006 for XMM1747 and July 23, 2006 for SAX1806 using the 
wide field infrared camera (WFCAM) composed of  four Rockwell Hawaii-II 2048 $\times$ 2048 18 micron pixel array detectors, 
with a pixel scale of 0\farcs4.  We determined  the zero point for our observations using 25 or more well-isolated 
and well-detected stars.
Because we calibrated our $K_\mathrm{s}$ band observations with respect to the UKIRT $K$ band  observations,  the magnitudes are discussed in $K$ band in the further discussions. 
The observed magnitude of both sources  from our observations are listed in Table 1.
 
We  also utilise data from the {\it  Spitzer} Space Telescope's \citep{Werner2004} Galactic Legacy Infrared Mid-Plane survey 
Extraordinaire (GLIMPSE; \citealt{Benjamin2003}) for XMM1747.  These data are also listed in Table 1. 

\section{Results}
\subsection{XMMU J174716.1--281048}

Our Magellan observations of the source in $J$, $H$ and $K_\mathrm{s}$ bands  were taken during the recent X-ray outburst which lasted $\sim$ 12 years from March 2003 to March 2015. 
Figure 1 shows the Magellan $K_\mathrm{s}$ band image of XMM1747 where the {\it Chandra} X-ray position of the source is marked with its error circle. No other X-ray star was detected in the {\it Chandra} image, therefore 
we could not further refine the position of 
the source through astrometry.  We detect two NIR stars consistent with the  {\it Chandra} 
error circle of the source at positions RA, Dec = 17$^\mathrm{h}$ 47$^\mathrm{m}$ 16$\fs$20,  
--28$^\circ$  10$^{\prime}$  47$\farcs$62  and 
17$^\mathrm{h}$ 47$^\mathrm{m}$ 16$\fs$16, 
--28$^\circ$  10$^{\prime}$  48$\farcs$72 
with a positional uncertainty of 0$\farcs$03 and are marked as C1, C2  in Figure 1.  
The observed magnitude of both stars in all three filters are listed in Table 1.

\begin{figure*}
\centering
\medskip
\includegraphics[height= 7cm]{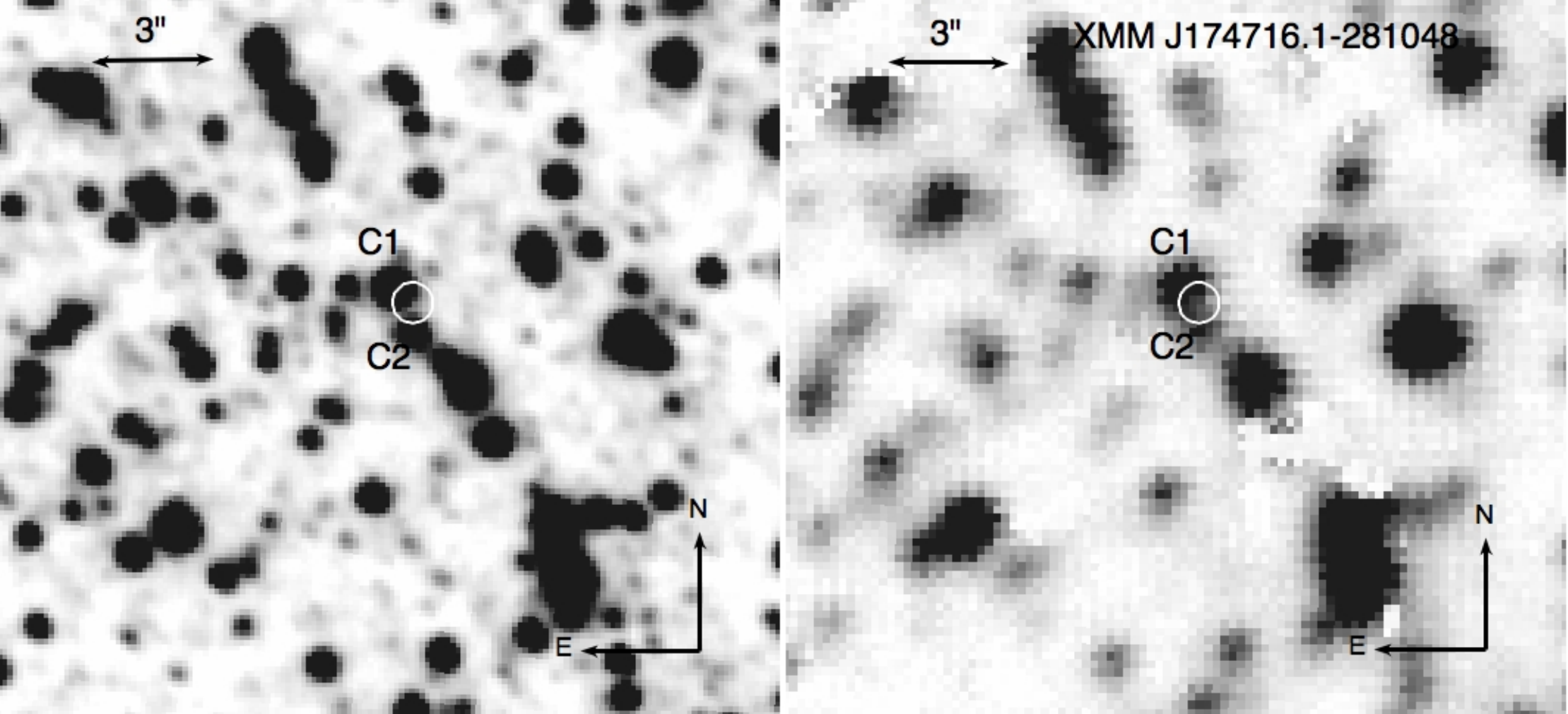}
\caption{Left : Magellan $K\mathrm{s}$ band image of XMM1747. Right : UKIRT $K$ band image of XMM1747.  In both images the white circle represents the Chandra X-ray position, with an error radius of 0\farcs5. 
The two candidate near-infrared counterparts are marked as C1 and C2. The white pixels in the UKIRT image might 
be  defects present in the CCD. }
\end{figure*}

The observations of the XMMJ1747 source field were also taken during the 
UKIDSS Galactic plane survey \citep{Lucas2008}  on July 18, 2006 in $J$, $H$ and $K$ bands.
Both stars are visible  in the $H$ and $K$ bands, see Figure 1, for the UKIRT $K$ band observations.  
However only star C1 is detected in the UKIRT J band observations. The magnitude of star C1 in
 $J$, $H$ and $K$ band is listed 
in the WFCAM science archive\footnote{http://surveys.roe.ac.uk/wsa/} of the UKIDSS surveys but not of  C2
which is likely because both stars are blended 
partially in these observations and C2 is detected much fainter than C1 during these observations.  
For a comparison, the  $K$ band  image of the source from both these observations are shown in Figure 1.
To measure the observed magnitude of  both stars C1 and C2 in the UKIRT images, 
we did photometry of these observations and calibrated with respect to the other stars in the field whose 
magnitude was present in the catalog.  
The magnitudes thus obtained in $J$ and $H$ and $K$ bands for both stars C1 and C2 are listed in Table 1. Because C2 was not detected in  J band image of the UKIRT observations, an upper limit on that position is calculated and  also listed in Table 1.

Both  the Magellan and UKIRT observations of the source  were 
taken (in 2008 and 2006 respectively), when the source was in outburst.  
During the two epochs, the magnitudes of both stars C1 and C2 are consistent with each other within the  errors 
(see Table 1). The source was also
observed using ANDICAM instrument on the SMARTS telescope with $H$ band magnitude of 15.3 $\pm$ 0.1 mag in 2007 
\citep{Degenaar2007ATel1136} which is consistent  with the $H$ band magnitude of  the brighter star C1. This
shows that the flux during these observations were dominated by that of C1.

We also looked at the color-color and color-magnitude 
diagrams of the  field using the  UKIDSS Galactic plane observations, shown in Figure 2.  
The two flux measurements of both candidate counterparts are also marked in the same Figure.  
Both candidate counterparts do not show any signs of reddening during the two observations 
and point towards being a part of the field population.

The spectral energy distribution (SED) of both candidate counterparts is also investigated.
Among the two candidates, only star C1 is listed in the GLIMPSE catalog as  SSTGLMA G000.8344+00.0834 and 
is detected at 3.6$\micron$ and 4.5$\micron$ with  12.20 $\pm$ 0.09 and 12.02 $\pm$ 0.11 mag, respectively. 
The Magellan observations allowed us to measure the magnitudes of both candidate counterparts with better confidence than the UKIRT 
observations. Therefore, to investigate the SED, we have used magnitudes from the Magellan observations. 
The  observed SED of  both stars C1 and C2 is shown in Figure 3. 
We estimated the extinction in the direction of the source in $J$, $H$ and $K$ bands 
using the relation N$_H$/Av=2.21 $\times$ 10$^{21}$ \citep{Guver2009}  and 
following \citet{Fitzpatrick1999}. The N$_\mathrm{H}$ of the source measured from the X-ray spectral fit  
is 8.9 $\times$ 10$^{22}$ cm$^{-2}$, which is  $\sim$ 8 times higher than the Galactic N$_\mathrm{H}$, 1.35 $\times$ 10$^{22}$ cm$^{-2}$ as measured by \citet{Dickey1990} in the direction of the source
using the Lyman-$\alpha$ and 21-cm lines. It is therefore possible that some of the absorption could be local to the X-ray source. 
For completeness,  we measured extinction in different bands using both N$_\mathrm{H}$ (henceforth A$^{\mathrm s}$ - 
extinction calculated using  N$_\mathrm{H}$ from the X-ray spectral fit, A$^{g}$ - 
extinction calculated using   N$_\mathrm{H}$ as measured by \citet{Dickey1990} in the direction of source). 
The extinction corrected magnitude of both stars from the Magellan and {\it Spitzer} observations are listed in Table 2. 
Here we assumed   that the NIR star is  experiencing the same N$_\mathrm{H}$ as that of X-ray source. However, if it is different,
then our results can change significantly.

A small accretion disk in the source would suggest that the optical/NIR radiation  is dominated by the 
companion star. If true, then we can find out  the spectral type of the companion star using a 
black-body fit to the SED. 
However the small range of wavelength covered during our observations and the large 
uncertainties in the extinction does not allow to fit the observed fluxes for realistic parameters of a star. 
Assuming that all  the NIR emission is  from the 
companion star, we estimated the distance to different type of stars (main-sequence, giants and supergiants; \citealt
{cox2000}) using the given de-reddened 
$K$ band flux and a black-body model. The $K$ band flux of star C1 corrected from A$^{s}$ suggest a star earlier than 
O5  or a supergiant of B-type 
while the one corrected from  A$^{g}$ suggest an early B-type main-sequence star, respectively at a distance of 8 kpc. 
Similarly, the $K$ band flux of star C2 from Magellan observations
corrected from both A$^{s}$ and A$^{g}$ suggest that it could be 
a star of spectral type O and late B-type main-sequence 
star, respectively. 
We also estimated the extinction and distance to different stars \citep{cox2000} using the observed colors of both candidate counterparts from both their 
Magellan and UKIRT observations and the results are consistent with the one using the black-body model, as shown in Figure 3. 

\begin{figure*}
\centering
\medskip
\includegraphics[width=15cm, height= 13cm]{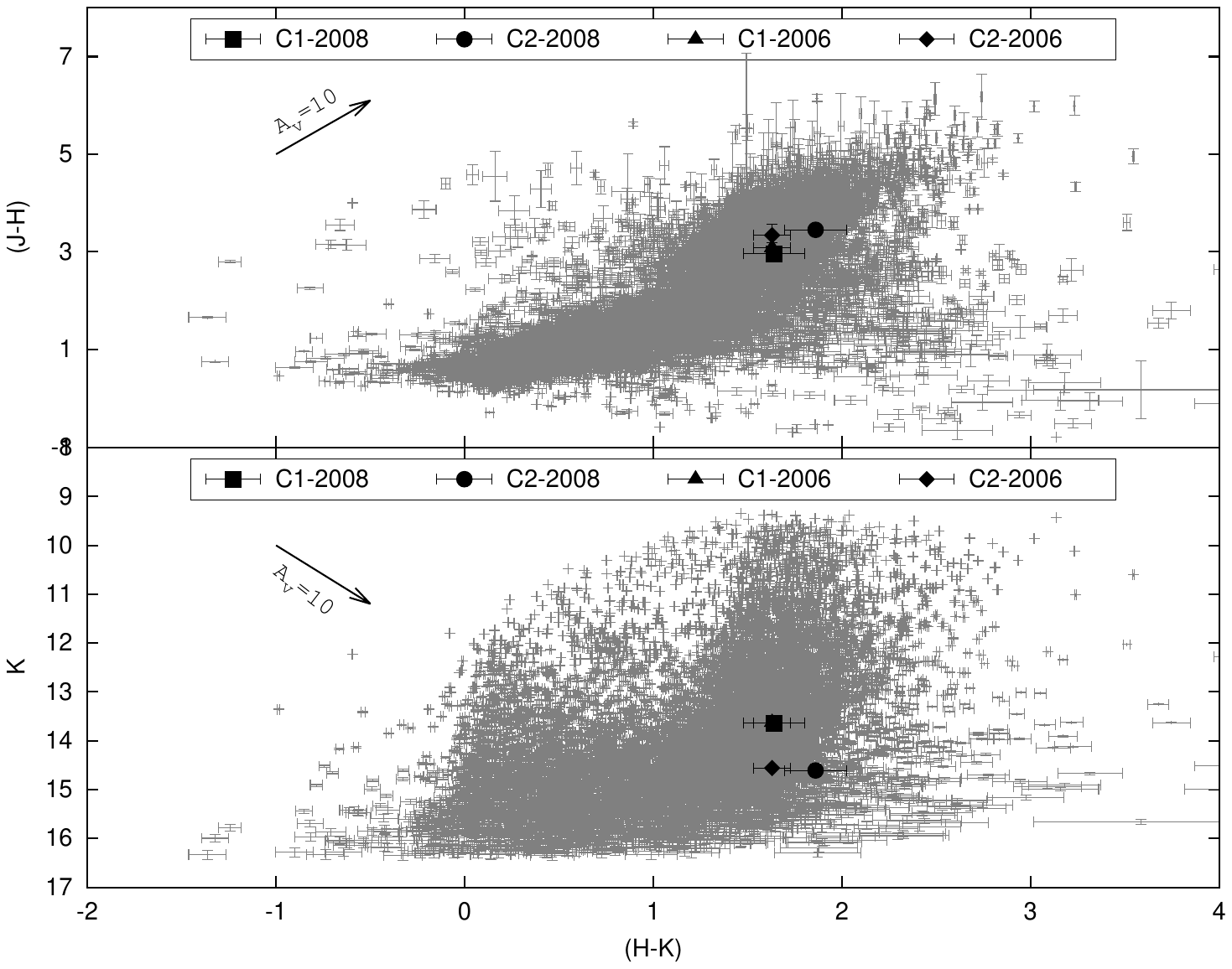}
\caption{Color-color and color-magnitude diagram of the source field XMM1747 from the 
UKIDSS Galactic plane survey observations. The arrow represents the extinction, Av=10 mag. Both Magellan (2008) and UKIRT (2006)
observations of candidate counterparts C1 and C2 are also marked. }
\end{figure*}

\begin{figure*}
\centering
\medskip
\includegraphics[height= 5.5cm]{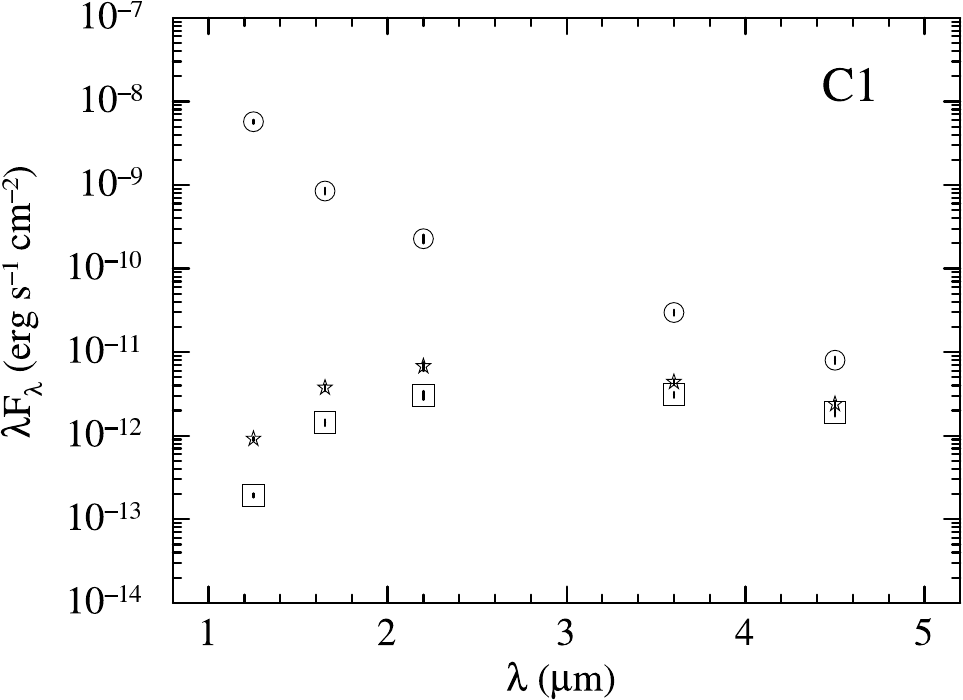}
\includegraphics[height= 5.5cm]{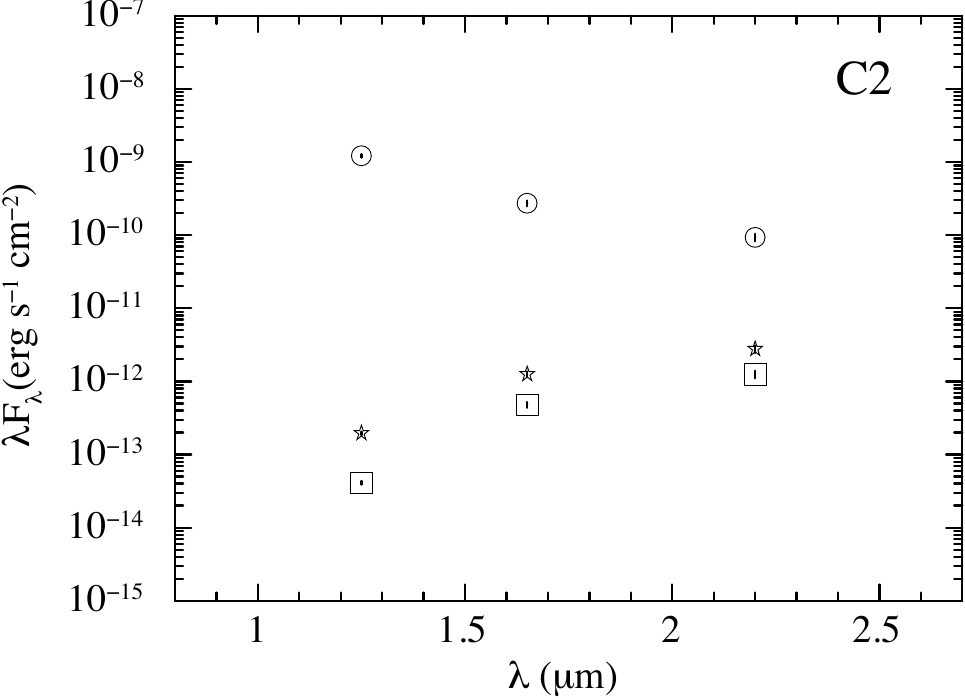}
\caption{The spectral energy distribution of  candidate counterparts C1 and C2 of XMM1747. 
The \texttt{squares} represent the observed fluxes from the Magellan and {\it Spitzer} observations, \texttt{circles} represent the de-redenned fluxes corrected from the extinction 
calculated using the N$_\mathrm{H}$ obtained from the X-ray spectral fit 
and \texttt{stars}  represent the  de-redenned fluxes, corrected from the extinction 
calculated using the Galactic N$_\mathrm{H}$ as given by \citet{Dickey1990}}
\end{figure*}

\begin{table*}
\centering
\caption{Observed near-infrared magnitudes of XMM1747 and SAX1806 from the Magellan and WIYN observations respectively. The UKIRT and  {\it Spitzer}/GLIMPSE magnitudes of the source are also listed. * on the UKIRT J band image represents the upper limit. }
\medskip
\begin{tabular}{lccccccc}
\hline
\hline
Telescope (Date) & Source name			& 	&& Filters \\ 
 & &  \emph{J} & \emph{H} & $K$  & 3.6 $\micron$ & 4.5 $\micron$\\ \hline
Magellan (May 25, 2008) & XMM1747 -  C1 	&  18.24 $\pm$ 0.06  	& 	15.28 $\pm$ 0.10      &  	13.64 $\pm$ 0.13  \\
& XMM1747 - C2 						&  19.92 $\pm$ 0.06     	&   	16.47 $\pm$ 0.10 	&  	14.61 $\pm$ 0.13 \\ \hline
UKIRT (July 18, 2006) & XMM1747 - C1 		&  18.34 $\pm$ 0.06 		&	15.26 $\pm$  0.09 	& 	13.63 $\pm$ 0.04 \\ 
& XMM1747 - C2  						&  19.9$^{*}$  		&    	16.19 $\pm$  0.09    	&   	14.56 $\pm$ 0.04 \\ \hline
Spitzer & XMM1747 - C1  &&&& 12.20 $\pm$ 0.09 & 12.02 $\pm$ 0.11 \\
\hline
 WIYN (March 23, 2011)  & SAX1806 & - & - & 17.25 $\pm$ 0.03 & - & - & \\  \hline \hline
\end{tabular}
\end{table*}

\begin{table*}
\centering
\caption{Extinction corrected magnitudes of XMM1747 and SAX1806 from the Magellan, WIYN and {\it Spitzer}/GLIMPSE observations.
The extinction in different bands are calculated using the N$_\mathrm{H}$ from the X-ray spectral fit (represented by superscript `s') 
and the Galactic N$_\mathrm{H}$  (represented by superscript `g') as measured by \citet{Dickey1990}. }
\medskip
\begin{tabular}{lcccccccc}
\hline
\hline
Telescope & Object			& 	N$_\mathrm{H}$  &    \emph{J} & $H$ & $K$  & 3.6$\micron$ & 4.5$\micron$\\ 
			& 	&  ($\times$ 10$^{22}$ cm$^{-2}$) &  (mag) & (mag) & (mag)		& (mag)	& (mag)\\ \hline
Magellan 		& XMM1747 -C1 		&  8.9$^{s}$	&		7.06 	$\pm$ 0.06  		& 	8.35  $\pm$ 0.10      	&  8.96  		$\pm$ 0.13 \\
			& XMM1747 -C2 		&  8.9$^{s}$	&		8.74   	$\pm$ 0.06    	&	9.58 $\pm$ 0.10 		&  9.93	 	$\pm$ 0.13 \\  
			& XMM1747 -C1 		&  1.4$^{g}$	&		16.55 	$\pm$ 0.06  		& 	14.24 $\pm$ 0.10      	&  12.78		$\pm$ 0.13  \\
			& XMM1747 -C2 		& 1.4$^{g}$	&		18.23   	$\pm$ 0.06    	&	15.43 $\pm$ 0.10 	&  13.75	 	$\pm$ 0.13     \\ \hline
{\it Spitzer} 	& XMM1747 - C1	 	& 8.9$^{s}$  & & & & 9.74  $\pm$ 0.09 	& 10.45 $\pm$ 0.11 \\
			& XMM1747 - C2 	& 1.4$^{g}$ &  & &  & 11.82  $\pm$ 0.09 	& 11.77  $\pm$ 0.11 \\ \hline 
WIYN 		& SAX1806 			& 5.6$^{s}$  & - & - &  14.31 $\pm$ 0.03 & - & -					\\ 
			& SAX1806 			& 1.2$^{g}$ & - & - &  16.62 $\pm$ 0.03 & - & - \\
\hline
\end{tabular}
\end{table*}

\subsection{SAX J1806.5--2215}

\begin{figure*}
\centering
\medskip
\includegraphics[height= 7cm]{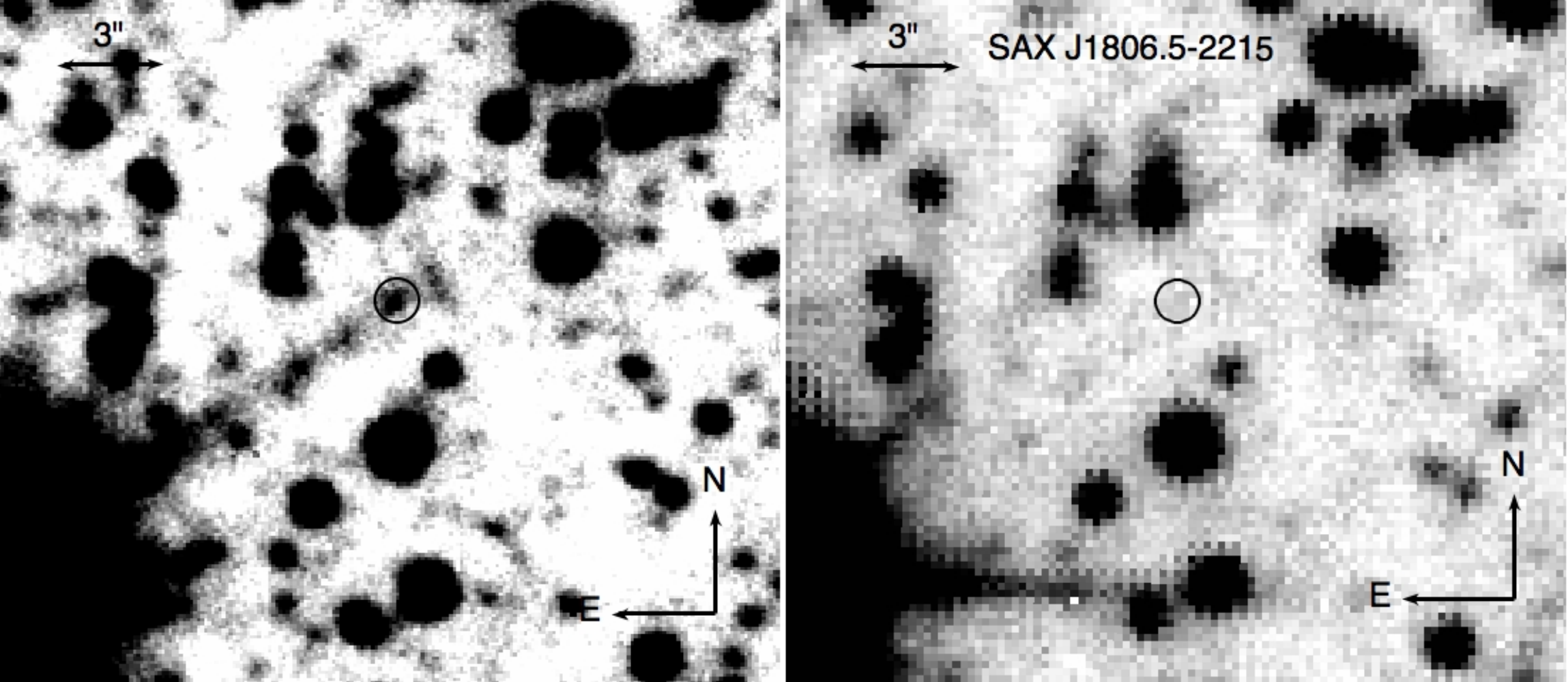}
\caption{Left : $K_\mathrm{s}$ band image of the SAX1806 taken using the WHIRC instrument on the 3.5-meter WIYN telescope. Right : $K$ band image of SAX1806 from the UKIDSS Galactic plane survey. }
\label{saxj1806}
\end{figure*}

We observed SAX1806 in the $K_\mathrm{s}$ waveband on March 23, 2011 during the ongoing X-ray outburst of the source, shown in Figure 4. 
The best measured {\it Chandra} X-ray position of the source is marked in the figure with the corresponding 0$\farcs$6 error 
circle. In our observations, we detected
one NIR star at RA, Dec (J2000) = 18$^\mathrm{h}$ 06$^\mathrm{m}$ 32$\fs$17, --22$^\circ$  14$^{\prime}$  17$\farcs$32, with an 
error circle of 0$\farcs$03 and 0$\farcs$05 in X and Y coordinates, 
consistent with the {\it Chandra} position of the source with a $K$ magnitude of 17.25 $\pm$ 0.03 mag.  We note that this 
magnitude differs by 0.4 mag as compared to the previous reporting of our observations \citep{Kaur2011ATel3268} 
and we believe this could be related to the calibration of the observations 
which was done with respect to the 2MASS during the previous 
analysis but with respect to the UKIRT observations in the analysis reported in this paper. Also it could be due to the zero-point calculations which are better 
constrained from the UKIRT observations. 

The source field was also observed during quiescence on July 23, 2006 in the NIR  $K$ band 
using the UKIRT telescope as a part of  the UKIDSS  Galactic plane survey but was not detected 
during these observations. 
The self calibration of the UKIRT $K$ band image gave an upper limit of 18.2 mag. This suggests that the 
 detected NIR star was brighter 
by almost a magnitude during the WIYN observations taken during the outburst. 

The N$_{\mathrm{H}}$ of the source from the X-ray spectral fit  is 4 times larger than the 
Galactic N$_{\mathrm{H}}$  measured in the direction
of the source.  Similar to XMM1747, the extinction is measured using both N$_{\mathrm{H}}$
and the extinction corrected magnitudes thus obtained are
listed in Table 2.  Assuming that the quiescence magnitude of the source is 18.2 mag or fainter in $K$ band, 
we estimate the companion star to be of spectral type K or later.

\section{Discussion}

In this paper, we aimed to find  possible NIR counterparts of two faint quasi-persistent X-ray binaries,
XMM1747 and SAX1806 that undergo
very long ($>$ years) X-ray outbursts. We obtained observations of XMM1747 in 2008 during its most recent outburst which lasted from March 2003 to March 2015, using the PANIC instrument on 
Magellan telescope and of  SAX1806 in March 2011 
using the WHIRC instrument on 3.5-meter WIYN telescope during its ongoing X-ray outburst which started in February 2011. 

We identified two NIR stars consistent with the {\it Chandra} error circle of XMM1747, marked as C1 and C2 in Figure 1. We compared
our observations taken in 2008 to UKIRT observations of the source obtained in 2006, both during the 
same outburst of the source. During both observations, 
stars C1 and C2 were detected with  similar magnitudes. 
Our color-color diagram analysis showed that both stars suffered almost the same 
extinction as most of the other stars in the field during both observations.
Assuming that all the NIR emission is from the companion star, the $K$ band flux of star C1 suggest 
a  high mass star of spectral type O or a supergiant of B-type to be the possible counterpart at a distance of 8 kpc.
Similarly, the $K$ band magnitude of star C2 suggest a high mass star O or late B-type 
from the Magellan and UKIRT observations respectively at a distance of 8 kpc.

None of  the sources which showed type-I X-ray bursts 
have ever been associated with a high-mass companion. 
Wind accretion from a high-mass counterpart can be a possibility, and if true, XMM1747 
would be the first X-ray burster from a high mass X-ray binary. 
However, the fact that XMM1747 is lying in a very crowded field in the NIR bands, which is also 
clear by the detection of  two stars in the {\it Chandra} error circle of the source, shows that the
chance coincidence of  detection of a foreground star is  high. Also the color-color diagram (Figure 2) 
of  both stars C1 and C2 indicate they are likely a part of the field population. 
Due to the high extinction towards the source, we cannot exclude that the 
real counterpart might not have been detected.  From the Magellan observations,  we measured the  
limiting magnitude of 20.0 mag in the $K$ band. 
Hence if the real counterpart was indeed not detected during our observations, 
then the observed $K$ band magnitude of the 
counterpart would be 20.0 mag or fainter, which indicates a  star of spectral type later than K, 
which is not unexpected for LMXBs. 
New X-ray observations have shown
that the source is back to quiescence and  follow-up NIR observations could give more insight into the NIR counterpart.

For SAX1806, we detected a NIR star of magnitude 17.25 $\pm$ 0.03 mag in the $K$ band during the X-ray outburst, 
consistent with its {\it Chandra} error circle, which makes it a very likely counterpart. 
This star was not detected during the UKIRT observations taken during the quiescence state of the source  suggesting a magnitude of 18.2 mag or fainter in $K$ band at that time. 
The increase in brightness by at least almost a magnitude during the X-ray outburst as compared to the quiescence suggest the 
detected NIR star is indeed  the likely counterpart of SAX1806. 
If true, and assuming the NIR magnitude of the companion star fainter than 18.2 mag, the companion 
star of SAX1806 would be close to a star of spectral type K or later.

As discussed above, it is very likely that the NIR counterpart of XMM1747 is a main sequence companion star K or later type and SAX1806 also suggest a counterpart being of spectral type K or later.  
If true, then it might be possible that indeed in those sources the faintness can be explained by the inhibition of matter by the magnetic field of the neutron stars as proposed by  \cite{Heinke2009}, \cite{DAngelo2012}
and \cite{Heinke2015}. Note however, that this model needs still to be confirmed.
However if the companion star is fainter than the magnitudes suggested (which is possible for both the sources) then they  could still be ultra compact X-ray binaries. Deeper OIR observations have to be performed 
in different wavebands to confirm their spectral type through the 
spectral energy distribution or if possible through the direct spectroscopic observations. 
In normal LMXBs, the OIR spectrum usually shows a number of 
double profile emission lines of H and He \citep{Kaur2012MAXI} during the outburst, indicating the presence of an accretion 
disk while in the quiescence state, it is dominated by the spectral 
signatures of the companion star. In ultra compact X-ray binaries, the accretion disk might still 
dominate the quiescence state. 

\section*{Acknowledgements} 
The United Kingdom Infrared Telescope is operated 
by the Joint Astronomy Centre on behalf of the Science 
and Technology Facilities Council of the U.K.
RK and RW acknowledge support from the ERC starting grant awarded to RW. 
RW also acknowledges support by a NWO top grant, module 1.
ND acknowledges support via an EU Marie Curie Intra-European fellowship 
under contract no. FP-PEOPLE-2013-IEF-627148


\begin{thebibliography}{58}
\expandafter\ifx\csname natexlab\endcsname\relax\def\natexlab#1{#1}\fi
\expandafter\ifx\csname url\endcsname\relax
  \def\url#1{{\tt #1}}\fi
\expandafter\ifx\csname urlprefix\endcsname\relax\def\urlprefix{URL }\fi

\bibitem[{Altamirano} et~al.(2011)]{Altamirano2011}
{Altamirano} D., {Kaur} R., {Degenaar} N. et~al.
\newblock The Astronomer's Telegram, 3193, 1 (2011).

\bibitem[{Armas Padilla} et~al.(2011)]{Armas2011}
{Armas Padilla} M., {Degenaar} N., {Patruno} A. et~al.
\newblock \mnras, 417, 659 (2011).

\bibitem[{Armas Padilla} et~al.(2014)]{Armas2014}
{Armas Padilla} M., {Wijnands} R., {Altamirano} D. et~al.
\newblock \mnras, 439, 3908 (2014).


\bibitem[{Arnason} et~al.(2015)]{Arnason2015}
{Arnason} R.M., {Sivakoff} G.R., {Heinke} C.O., {Cohn} H.N. and {Lugger} P.M.
\newblock \apj, 807, 52 (2015).

\bibitem[{Benjamin} et~al.(2003)]{Benjamin2003}
{Benjamin} R.A., {Churchwell} E., {Babler} B.L. et~al.
\newblock \pasp, 115, 953 (2003).

\bibitem[{Brandt} et~al.(2006)]{Brandt2006ATel970}
{Brandt} S., {Budtz-Jorgensen} C., {Chenevez} J. et~al.
\newblock The Astronomer's Telegram, 970, 1 (2006).

\bibitem[{Campana}(2009{\natexlab{a}})]{Campana2009ApJ}
{Campana} S.
\newblock \apj, 699, 1144 (2009{\natexlab{a}}).

\bibitem[{Campana}(2009{\natexlab{b}})]{Campana2009}
{Campana} S.
\newblock \apj, 699, 1144 (2009{\natexlab{b}}).

\bibitem[{Chakrabarty} et~al.(2011)]{Chakrabarty2011ATel3218}
{Chakrabarty} D., {Jonker} P. and {Markwardt} C.B.
\newblock The Astronomer's Telegram, 3218, 1 (2011).

\bibitem[{Cornelisse} et~al.(2002{\natexlab{a}})]{Cornelisse2002}
{Cornelisse} R., {Verbunt} F., {in't Zand} J.J.M., {Kuulkers} E. and {Heise} J.
\newblock \aap, 392, 931 (2002{\natexlab{a}}).

\bibitem[{Cornelisse} et~al.(2002{\natexlab{b}})]{Cornelisse2002AA392}
{Cornelisse} R., {Verbunt} F., {in't Zand} J.J.M. et~al.
\newblock \aap, 392, 885 (2002{\natexlab{b}}).

\bibitem[{Corral-Santana} et~al.(2013)]{Corral2013}
{Corral-Santana} J. M., {Casares} J., {Mu{\~n}oz-Darias} T. et~al.
\newblock Science, 339, 1048C (2013).


\bibitem[{Cox}(2000)]{cox2000}
{Cox} A.N.
\newblock {Allen's astrophysical quantities} (2000).

\bibitem[{D'Angelo} and {Spruit}(2012)]{DAngelo2012}
{D'Angelo} C.R. and {Spruit} H.C.
\newblock \mnras, 420, 416 (2012).

\bibitem[{Degenaar} et~al.(2011{\natexlab{a}})]{Degenaar2011ATel3202}
{Degenaar} N., {Altamirano} D., {Padilla} M.A. et~al.
\newblock The Astronomer's Telegram, 3202, 1 (2011{\natexlab{a}}).

\bibitem[{Degenaar} et~al.(2010)]{Degenaar2010MNRAS}
{Degenaar} N., {Jonker} P.G., {Torres} M.A.P. et~al.
\newblock \mnras, 404, 1591 (2010).

\bibitem[{Degenaar} et~al.(2007)]{Degenaar2007ATel1136}
{Degenaar} N., {Krauss} M., {Maitra} D. et~al.
\newblock The Astronomer's Telegram, 1136, 1 (2007).

\bibitem[{Degenaar} and {Wijnands}(2007)]{Degenaar2007ATel1078}
{Degenaar} N. and {Wijnands} R.
\newblock The Astronomer's Telegram, 1078, 1 (2007).

\bibitem[{Degenaar} and {Wijnands}(2010)]{Degenaar2010AA}
{Degenaar} N. and {Wijnands} R.
\newblock \aap, 524, A69 (2010).

\bibitem[{Degenaar} et~al.(2011{\natexlab{b}})]{Degenaar2011}
{Degenaar} N., {Wijnands} R. and {Kaur} R.
\newblock \mnras, 414, L104 (2011{\natexlab{b}}).

\bibitem[{Del Santo} et~al.(2007{\natexlab{a}})]{DelSanto2007ATel1207}
{Del Santo} M., {Chenevez} J., {Brandt} S., {Bazzano} A. and {Ubertini} P.
\newblock The Astronomer's Telegram, 1207, 1 (2007{\natexlab{a}}).

\bibitem[{Del Santo} et~al.(2012{\natexlab{a}})]{DelSanto2012ATel4017}
{Del Santo} M., {Romano} P., {Ferrigno} C. et~al.
\newblock The Astronomer's Telegram, 4017, 1 (2012{\natexlab{a}}).

\bibitem[{Del Santo} et~al.(2010)]{DelSanto2010ATel2624}
{Del Santo} M., {Romano} P. and {Sidoli} L.
\newblock The Astronomer's Telegram, 2624, 1 (2010).

\bibitem[{Del Santo} et~al.(2011)]{DelSanto2011ATel3471}
{Del Santo} M., {Romano} P. and {Sidoli} L.
\newblock The Astronomer's Telegram, 3471, 1 (2011).

\bibitem[{Del Santo} et~al.(2012{\natexlab{b}})]{DelSanto2012ATel4099}
{Del Santo} M., {Romano} P. and {Sidoli} L.
\newblock The Astronomer's Telegram, 4099, 1 (2012{\natexlab{b}}).

\bibitem[{Del Santo} et~al.(2015)]{DelSanto2015ATel7293}
{Del Santo} M., {Romano} P. and {Sidoli} L.
\newblock The Astronomer's Telegram, 7293, 1 (2015).

\bibitem[{Del Santo} et~al.(2009)]{DelSanto2009ATel2050}
{Del Santo} M., {Romano} P., {Sidoli} L. and {Bazzano} A.
\newblock The Astronomer's Telegram, 2050, 1 (2009).

\bibitem[{Del Santo} et~al.(2007{\natexlab{b}})]{DelSanto2007}
{Del Santo} M., {Sidoli} L., {Mereghetti} S. et~al.
\newblock \aap, 468, L17 (2007{\natexlab{b}}).

\bibitem[{Dickey} and {Lockman}(1990)]{Dickey1990}
{Dickey} J.M. and {Lockman} F.J.
\newblock \araa, 28, 215 (1990).

\bibitem[{Fitzpatrick}(1999)]{Fitzpatrick1999}
{Fitzpatrick} E.L.
\newblock \pasp, 111, 63 (1999).

\bibitem[{G{\"u}ver} and {{\"O}zel}(2009)]{Guver2009}
{G{\"u}ver} T. and {{\"O}zel} F.
\newblock \mnras, 400, 2050 (2009).


\bibitem[{Hameury} and {Lasota}(2016)]{Hameury2016}
{Hameury} J, M and {Lasota} J, P.
\newblock arXiv 1607.06394 (2016).



\bibitem[{Heinke} et~al.(2009)]{Heinke2009}
{Heinke} C.O., {Cohn} H. N., and {Lugger} P. M. 
\newblock \mnras, 447, 3034 (2015).

\bibitem[{Heinke} et~al.(2015)]{Heinke2015}
{Heinke} C.O., {Bahramian} A., {Degenaar} N. and {Wijnands} R.
\newblock \mnras, 447, 3034 (2015).

\bibitem[{Heinke} et~al.(2009)]{Heinke2009}
{Heinke} C.O., {Cohn} H.N. and {Lugger} P.M.
\newblock \apj, 692, 584 (2009).

\bibitem[{in't Zand} et~al.(2005)]{IntZand2005}
{in't Zand} J.J.M., {Cumming} A., {van der Sluys} M.V., {Verbunt} F. and {Pols}
  O.R.
\newblock \aap, 441, 675 (2005).

\bibitem[{In't Zand} et~al.(1999)]{In'tzand1999}
{In't Zand} J.J.M., {Heise} J., {Muller} J.M. et~al.
\newblock Nuclear Physics B Proceedings Supplements, 69, 228 (1999).

\bibitem[{in't Zand} et~al.(2009)]{intZand2009}
{in't Zand} J.J.M., {Jonker} P.G., {Bassa} C.G., {Markwardt} C.B. and {Levine}
  A.M.
\newblock \aap, 506, 857 (2009).

\bibitem[{in't Zand} et~al.(2007)]{IntZand2007}
{in't Zand} J.J.M., {Jonker} P.G. and {Markwardt} C.B.
\newblock \aap, 465, 953 (2007).

\bibitem[{Kaur} et~al.(2011{\natexlab{a}})]{Kaur2011ATel3695}
{Kaur} R., {Heinke} C., {Kotulla} R. et~al.
\newblock The Astronomer's Telegram, 3695, 1 (2011{\natexlab{a}}).

\bibitem[{Kaur} et~al.(2012{\natexlab{a}})]{Kaur2012MAXI}
{Kaur} R., {Kaper} L., {Ellerbroek} L.E. et~al.
\newblock \apjl, 746, L23 (2012{\natexlab{a}}).

\bibitem[{Kaur} et~al.(2011{\natexlab{b}})]{Kaur2011ATel3268}
{Kaur} R., {Kotulla} R., {Degenaar} N., {Wijnands} R. and {Kaplan} D.
\newblock The Astronomer's Telegram, 3268, 1 (2011{\natexlab{b}}).

\bibitem[{Kaur} et~al.(2012{\natexlab{b}})]{Kaur2012ATel3926}
{Kaur} R., {Wijnands} R., {Heinke} C. and {Degenaar} N.
\newblock The Astronomer's Telegram, 3926, 1 (2012{\natexlab{b}}).

\bibitem[{Kaur} et~al.(2009)]{Kaur2009}
{Kaur} R., {Wijnands} R., {Patruno} A. et~al.
\newblock \mnras, 394, 1597 (2009).

\bibitem[{Kaur} et~al.(2010)]{Kaur2010}
{Kaur} R., {Wijnands} R., {Paul} B., {Patruno} A. and {Degenaar} N.
\newblock \mnras, 402, 2388 (2010).

\bibitem[{King} and {Wijnands}(2006)]{King2006}
{King} A.R. and {Wijnands} R.
\newblock \mnras, 366, L31 (2006).

\bibitem[{Lasota}(2001)]{Lasota2001}
{Lasota} J.P.
\newblock \nar, 45, 449 (2001).

\bibitem[{Lucas} et~al.(2008)]{Lucas2008}
{Lucas} P.W., {Hoare} M.G., {Longmore} A. et~al.
\newblock \mnras, 391, 136 (2008).

\bibitem[{Maccarone} and {Patruno}(2013)]{Maccarone2013}
{Maccarone} T.J. and {Patruno} A.
\newblock \mnras, 428, 1335 (2013).

\bibitem[{Martini} et~al.(2004)]{Martini2004}
{Martini} P., {Persson} S.E., {Murphy} D.C. et~al.
\newblock {Moorwood} A.F.M. and {Iye} M., (eds.) Ground-based Instrumentation
  for Astronomy, volume 5492 of Society of Photo-Optical Instrumentation
  Engineers (SPIE) Conference Series, 1653--1660 (2004).

\bibitem[{Meixner} et~al.(2010)]{Meixner2010}
{Meixner} M., {Smee} S., {Doering} R.L. et~al.
\newblock \pasp, 122, 451 (2010).

\bibitem[{Nelemans} and {Jonker}(2010)]{Nelemans2010}
{Nelemans} G. and {Jonker} P.G.
\newblock \nar, 54, 87 (2010).

\bibitem[{Pfahl} et~al.(2002)]{Pfahl2002ApJ}
{Pfahl} E., {Rappaport} S. and {Podsiadlowski} P.
\newblock \apjl, 571, L37 (2002).

\bibitem[{Rau} et~al.(2011)]{Rau2011}
{Rau} A., {Greiner} J. and {Filgas} R.
\newblock  The Astronomer's Telegram, 3140, 1 (2011). 

\bibitem[{Sidoli} et~al.(2004)]{Sidoli2004}
{Sidoli} L., {Bocchino} F., {Mereghetti} S. and {Bandiera} R.
\newblock Memorie della Societ Astronomica Italiana, 75, 507 (2004).

\bibitem[{Sidoli} and {Mereghetti}(2003)]{Sidoli2003ATel147}
{Sidoli} L. and {Mereghetti} S.
\newblock The Astronomer's Telegram, 147, 1 (2003).

\bibitem[{Sidoli} et~al.(2006)]{Sidoli2006}
{Sidoli} L., {Mereghetti} S., {Favata} F., {Oosterbroek} T. and {Parmar} A.N.
\newblock \aap, 456, 287 (2006).

\bibitem[{Sidoli} et~al.(2007)]{Sidoli2007ATel1174}
{Sidoli} L., {Romano} P., {Mereghetti} S. and {Del Santo} M.
\newblock The Astronomer's Telegram, 1174, 1 (2007).

\bibitem[{Skrutskie} et~al.(2006)]{Skrutskie2006}
{Skrutskie} M.F., {Cutri} R.M., {Stiening} R. et~al.
\newblock \aj, 131, 1163 (2006).

\bibitem[{Sguera} et~al.(2015)]{VSguera2015ATel8222}
{Sguera} V., {Sidoli} L., {Fiocchi} M., {Bazzano} A., {Ubertini} P., {Paizis} A., and {Kuulkers} E.
\newblock The Astronomer's Telegram, 8222, 1 (2015).

\bibitem[{Stetson} et~al.(1987)]{Stetson1987}
{Stetson} P. B.
\newblock PASP, 99, 191 (1987).


\bibitem[{Werner} et~al.(2004)]{Werner2004}
{Werner} M.W., {Roellig} T.L., {Low} F.J. et~al.
\newblock \apjs, 154, 1 (2004).

\bibitem[{Wijnands}(2006)]{Wijnands2006ATel972}
{Wijnands} R.
\newblock The Astronomer's Telegram, 972, 1 (2006).

\bibitem[{Wijnands} et~al.(2006)]{Wijnands2006AA}
{Wijnands} R., {in't Zand} J.J.M., {Rupen} M. et~al.
\newblock \aap, 449, 1117 (2006).

\bibitem[{Wijnands} et~al.(2009)]{Wijnands2009}
{Wijnands} R., {Rol} E., {Cackett} E., {Starling} R.L.C. and {Remillard} R.A.
\newblock \mnras, 393, 126 (2009).

\end{thebibliography}
\end{document}